\documentclass[sigconf,screen]{acmart}
\setcopyright{none}
\renewcommand\footnotetextcopyrightpermission[1]{}
\acmDOI{}\acmISBN{}\acmConference{}{}{}\acmYear{2026}\copyrightyear{2026}
\usepackage{booktabs,graphicx,amsmath,xspace}
\usepackage[ruled,vlined]{algorithm2e}
\usepackage{enumitem,xurl,longtable,array}
\makeatletter
\@ifundefined{hyxmp@parse@acmart}{}{%
  \patchcmd{\hyxmp@parse@acmart}{\begingroup}{\begingroup
    \let\@authornotemark\relax\let\@@authornotemark\@gobble}{}{}%
}
\makeatother
\title{Keeping Fuzzers on Duty: Repairing Fuzzing Build Breakages in Continuous Open-Source Fuzzing}
\author{Zhuang Zhao}
\authornote{Zhuang Zhao and Jiongchi Yu contributed equally to this work.}
\affiliation{\institution{Tianjin University}\city{Tianjin}\country{China}}
\email{zhaozhuang@tju.edu.cn}
\author{Jiongchi Yu}
\authornotemark[1]
\affiliation{\institution{Nanyang Technological University}\country{Singapore}}
\email{jiongchiyu@acm.org}
\author{Yao Zhang}
\authornote{Corresponding authors: Yao Zhang and Xiaohong Li.}
\affiliation{\institution{Tianjin University}\city{Tianjin}\country{China}}
\email{zzyy@tju.edu.cn}
\author{Dongge Liu}
\affiliation{\institution{Google}\country{Australia}}
\email{donnggeliu@google.com}
\author{Qiang Hu}
\affiliation{\institution{Tianjin University}\city{Tianjin}\country{China}}
\email{qianghu@tju.edu.cn}
\author{Xiaohong Li}
\authornotemark[2]
\affiliation{\institution{Tianjin University}\city{Tianjin}\country{China}}
\email{xiaohongli@tju.edu.cn}
\author{Yang Liu}
\affiliation{\institution{Nanyang Technological University}\country{Singapore}}
\email{yangliu@ntu.edu.sg}

\fancypagestyle{standardpagestyle}{\fancyhf{}\fancyfoot[C]{\thepage}\renewcommand{\headrulewidth}{0pt}}
\fancypagestyle{firstpagestyle}{\fancyhf{}\fancyfoot[C]{\thepage}\renewcommand{\headrulewidth}{0pt}}
\begin{document}
\begin{abstract}
Continuous fuzzing platforms such as OSS-Fuzz provide sustained security testing for open-source projects, but long-term build breakages reduce security coverage and waste infrastructure resources. In OSS-Fuzz, around $20\%$ of integrated projects stay in build-failure states, leaving their fuzzers effectively off duty. These failures expose a critical maintenance gap between platform maintainers, who lack project-specific context, and project maintainers, who often lack fuzzing and infrastructure expertise. Existing program-repair techniques cannot be directly applied as fuzz build breakages span evolving project and infrastructure artifacts. They require repeated infrastructure-aware diagnosis rather than localized code edits, making them difficult to repair successfully.

We present FuzzMedic, an automated agent for fuzz build repair. FuzzMedic performs evidence-guided diagnosis to identify repair opportunities across project and fuzzing-infrastructure artifacts. It then conducts repair exploration over a repair-state graph (RSG), which captures the evolving workspace and builds feedback throughout the repair process. By supporting branching and rollback, the RSG enables FuzzMedic to recover from unproductive repair paths during long-horizon system-level debugging. FuzzMedic further validates candidate patches to ensure that they restore executable fuzz targets rather than merely suppress build failures.

We evaluate FuzzMedic on $128$ reproducible OSS-Fuzz build failures extracted from $1,820$ real-world failures. FuzzMedic achieves a $48.44\%$ repair rate and outperforms three representative program-repair baselines. We further deploy FuzzMedic on real-world OSS-Fuzz failures, where it repairs $29$ real-world OSS-Fuzz cases that have been merged by upstream maintainers, including 3 long-standing failures that had remained unresolved for more than one year and 2 cases that had never built successfully.

\end{abstract}
\maketitle
\pagestyle{fancy}
\fancyhf{}
\fancyfoot[C]{\thepage}
\renewcommand{\headrulewidth}{0pt}
\section{Introduction}
Continuous fuzzing services keep testing open-source software as projects evolve. In OSS-Fuzz, project-specific integrations build instrumented targets inside containers and deliver executable fuzzers to the testing infrastructure~\cite{ossfuzz,serebryany2017oss,ossfuzz_docs}. When this build fails, the affected configuration cannot supply newly built fuzz targets. Repairing OSS-Fuzz build failures is therefore necessary to maintain continuous security testing, even when the project still builds successfully outside the fuzzing environment.

OSS-Fuzz-specific evidence establishes the maintenance problem. Nourry et al. analyze more than 1.2 million OSS-Fuzz build logs and characterize failures involving project code, dependencies, fuzzing environments, Docker environments, and external resources~\cite{nourry2025my}. Our collected cases similarly require reasoning about upstream revisions alongside OSS-Fuzz build scripts, toolchains, and sanitizer settings. The operational difficulty is the split between project-specific build knowledge and platform integration knowledge. This motivates repair assistance across both workspaces. It does not establish that every compilation error is intrinsically specific to fuzzing.

Existing build-repair systems address important parts of this problem. BuildMedic targets Maven dependency breakage, CXXCrafter targets general C/C++ software building, and CompileAgent automates repository-level compilation~\cite{macho2018automatically,yu2025cxxcrafter,hu2025compileagent}. Their reported results concern different tasks and datasets and are not directly comparable with OSS-Fuzz repair rates. Generic repository-repair systems also offer useful localization and search strategies~\cite{xia2024agentless,antoniades2024swe}. In our setting, however, the actionable change may lie outside the upstream repository, and repair must satisfy container, instrumentation, and fuzz-target validation constraints.

Three practical demands guide our design. First, a compiler or linker message identifies a symptom, while the responsible change may lie in another workspace or dependency. Second, failures before a runnable target exists provide build feedback rather than crash-triggering inputs. Third, patch attempts modify the workspace. Clearing an early error can expose a later error, while an unsuccessful edit can damage previously working stages. Repeated full builds make indiscriminate patch sampling costly. Section~\ref{sec:background} illustrates these demands with two repair trajectories.

We present FuzzMedic, implemented as collaborating agents using Google ADK. It connects Evidence-Constrained Root-Cause Localization (ECRCL), Failure-Guided Repair Knowledge (FGRK), Reflection Summary and Memory Curation (RSMC), and Hierarchical Stateful Rollback (HSR). The contribution is their integration around observable OSS-Fuzz build states. Localization constrains candidate changes, templates encode domain constraints, memory records validated outcomes, and rollback preserves recoverable workspaces.

We evaluate this system on 128 reproducible failures from 113 projects. The reported configuration repairs 62 cases (48.44\%), versus 27 (21.09\%) for the strongest native-budget baseline. Different native budgets limit causal interpretation of this gap. Supplementary comparisons use matched token/time caps or limited subsets. Separately, 29 generated patches were merged into upstream or OSS-Fuzz main branches after human verification and maintainer review. The workflow was also exercised through a pull request (PR) experiment in \textit{oss-fuzz-gen}. This demonstrates an experimental integration rather than production adoption.

The paper makes the following three contributions.
\begin{itemize}[leftmargin=*,nosep]
\item A reproducible OSS-Fuzz build-repair task and benchmark, with pinned failure contexts, explicit selection criteria, language distributions, and category-level outcomes.
\item An ADK-based multi-agent repair system combining cross-workspace evidence, repair knowledge, trajectory memory, and recoverable build-state exploration.
\item An evaluation of repair effectiveness and resource use. In collaboration with Google, we integrate FuzzMedic into an \textit{oss-fuzz-gen} PR experiment and produce 29 maintainer-merged patches across 28 projects.
\end{itemize}

\begin{figure}[t]
\centering
\includegraphics[width=0.8\columnwidth]{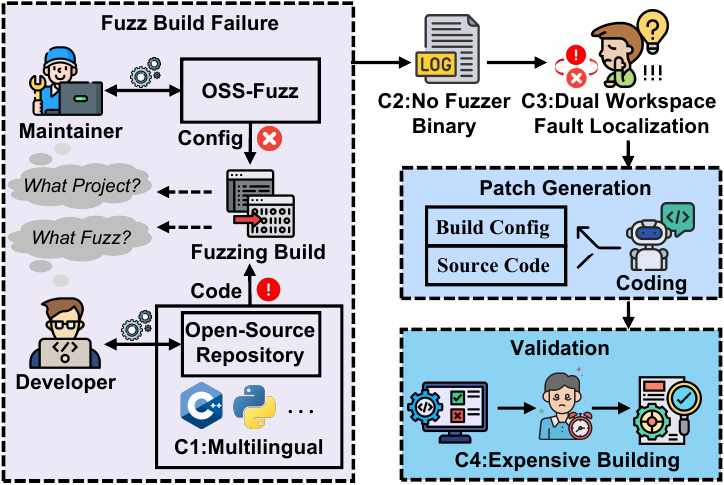}
\caption{Overview of fuzz build repair in OSS-Fuzz.}
\label{fig:fuzz-build-task}
\Description{OSS-Fuzz build repair uses failure logs to localize faults across upstream source and fuzzing configuration, generates patches, and validates them through rebuilding.}
\end{figure}

\section{Background}
\label{sec:background}


\subsection{Continuous Fuzzing Platforms}

Fuzzing searches for security and stability defects through generated inputs~\cite{manes2019art}. Continuous services automate instrumented builds, execution, and reporting. Developers supply targets, scripts, and sanitizer configurations. The platform builds and validates executable targets before fuzzing. A failure at this prerequisite prevents downstream feedback from newly built targets.

OSS-Fuzz~\cite{ossfuzz} is a representative centralized platform for open-source software. It improves the security and stability of common open-source software by combining modern fuzzing techniques with scalable distributed execution~\cite{serebryany2017oss}. Its workflow builds project container images, compiles fuzz targets under specified sanitizer, engine, and architecture settings, and validates the generated fuzzers before continuous fuzzing begins~\cite{ossfuzz_docs}.

ClusterFuzzLite~\cite{clusterfuzzlite} integrates fuzzing with continuous integration and continuous delivery (CI/CD). Code Intelligence provides CI Fuzz integration examples for CMake and JUnit~\cite{cifuzz}. Our evaluation is based solely on OSS-Fuzz.

\subsection{Automated Build Repair for Fuzzing}

Figure~\ref{fig:fuzz-build-task} summarizes the task, from dual-workspace fault localization to patch generation and build validation. BuildMedic repairs Maven dependency failures, while CXXCrafter automates general C/C++ builds~\cite{macho2018automatically,yu2025cxxcrafter}. OSS-Fuzz additionally requires compatible instrumentation, integration artifacts, and target validation.

Our focus is the repair of OSS-Fuzz fuzzing-build failures before a runnable target is available. Existing work on OSS-Fuzz and automated repair addresses adjacent tasks. Nourry et al.~\cite{nourry2025my} characterize and classify build failures but do not generate patches. Zhang et al.~\cite{zhang2024fixing} use OSS-Fuzz vulnerability reports to repair security bugs, assuming that a successful fuzzing campaign has already produced crash inputs. Kim et al.~\cite{kim2023automated} connect fuzzing and automated program repair (APR) at the patch-search level rather than the build-maintenance level. This gap motivates a repair system designed for OSS-Fuzz fuzzing build failures.

\subsection{Motivation Example}

To illustrate the additional constraints encountered in this setting, we analyze two OSS-Fuzz failures where Agentless, CompileAgent, and SWE-Search fail but FuzzMedic succeeds.

\noindent\textbf{Root-Cause Evidence in \textit{jimfs}.}
The build log reports a Java class-version mismatch. Error Prone was compiled for Java~21 with class file version~65.0, while the OSS-Fuzz configuration uses Java~17 with version~61.0. Several repairs appear plausible, including source edits, Java Development Kit (JDK) upgrades, and dependency downgrades. The identified cause is an upstream change from \textit{errorprone.version} 2.42.0 to 2.43.0, which introduces an annotation processor incompatible with this JDK profile. The baselines identify the symptom but miss the responsible commit. FuzzMedic localizes that commit and generates a compatible fix in one round.

\noindent\textbf{State-Aware Repair in \textit{toml\_edit}.}
The \textit{toml\_edit} failure requires sequential repair. A Docker \textit{WORKDIR /src} directive conflicts with the local checkout mechanism and prevents source retrieval. Fixing checkout exposes a Rust linker failure involving unresolved sanitizer-coverage symbols such as \textit{\_\_sancov\_gen\_*} under AddressSanitizer (ASan). The baselines alternate between these failures without preserving solved subproblems. FuzzMedic uses RSMC to retain round-level progress, FGRK to retrieve an OSS-Fuzz-specific \textit{WORKDIR} rule, and HSR to recover from regressive branches. This preserves checkout progress during linker repair.

These trajectories motivate combining evidence-guided localization, repair knowledge, and state recovery.

\section{Dataset Construction}
\label{sec:dataset}

Controlled repair evaluation requires reconstructing the historical build context, not merely collecting error messages. We therefore construct a curated benchmark for build failure repair and controlled evaluation. Following the taxonomy of Nourry et al.~\cite{nourry2025my}, we collect $1,820$ build failure logs from the OSS-Fuzz public build-status portal and log storage~\cite{ossfuzz_build_status}. We reproduce each candidate by restoring the recorded failure-time repositories and build environments, and retain $128$ reproducible cases from $113$ projects.

\subsection{Failure Collection}

The collected logs span December 29, 2022 to June 14, 2026. We collected $1,820$ build failure logs from the OSS-Fuzz public build-status portal~\cite{ossfuzz_build_status}, which links to archived failure logs. For each candidate, we restored the OSS-Fuzz repository, upstream project, and third-party dependencies to the recorded failure-time versions, re-executed the fuzz-build pipeline in Docker, and retained the case only when the reproduced failure matched the original error message and failure stage.

\subsection{Selection and Reproducibility}
Candidate evaluation records are archived build-failure logs, maintained separately from FGRK's historical failure-and-repair source records. Their collection windows coincide. Independence is established by FGRK's project and related-incident exclusions, not by assuming distinct collection dates.

We first merge records confirmed to arise from the same underlying breakage, removing 1,076 duplicates and retaining 744 representatives. Identical error text alone is insufficient to merge records. Unresolved relationships remain separate. We then require extractable logs and metadata and initially accessible version references and resources, excluding 256 records and retaining 488. A historical fix is not required for inclusion. Finally, we pin the OSS-Fuzz and upstream commits, base-image digest, and dependency commit identifiers, restore the historical environment, and replay the Docker build. Inclusion requires the original diagnostic and failure stage, confirmed manually. This excludes 360 records whose resources or environment cannot be recovered or whose original failure cannot be reproduced.

Table~\ref{tab:selection} reports mutually exclusive exclusions. Of the 1,692 records not retained, only 360 fail the historical restoration or replay. The remainder are duplicates or fail metadata checks. The resulting 128 cases form a controlled-replay benchmark, not a representative sample of all OSS-Fuzz build failures. Appendix~\ref{app:selection} details the audit requirements.
\begin{table}[t]
\centering\caption{Sequential benchmark selection. Exclusions are counted at the first applicable stage.}\label{tab:selection}
\begin{tabular}{lrr}\toprule Stage & Excluded & Retained\\\midrule
Collected logs &---&1,820\\
Incident deduplication &1,076&744\\
Metadata checks &256&488\\
Historical replay &360&128\\\midrule
Total excluded &1,692&\\\bottomrule
\end{tabular}
\end{table}

\subsection{Metadata Schema}

Each dataset entry is a structured record in YAML Ain't Markup Language (YAML) format. It contains the context needed to restore the environment and reproduce the failure (Table~\ref{tab:metadata}). Root-cause annotations and outcome labels are evaluation data, separate from the failure-time inputs available to the agent. Reference fixes and post-failure repair history are excluded from agent-accessible snapshots and tools. Labels and provenance remain audit-only.

\begin{table}[t]
\centering
\caption{Metadata schema for each dataset case.}
\label{tab:metadata}
\setlength{\tabcolsep}{4pt}
\resizebox{\columnwidth}{!}{%
\begin{tabular}{ll}
\toprule
\textbf{Field} & \textbf{Description} \\
\midrule
project                    & Project name \\
language                   & Primary programming language \\
error\_time                & Date of the observed build failure \\
oss-fuzz\_sha              & OSS-Fuzz commit identifier at failure time \\
fuzzing\_build\_error\_log & Link to the original OSS-Fuzz build log \\
software\_repo\_url        & Link to the upstream open-source repository \\
software\_sha              & Upstream project commit identifier at failure time \\
engine                     & Configured fuzzing engine \\
sanitizer                  & Sanitizer configuration, e.g., address \\
architecture               & Target processor architecture \\
base\_image\_digest        & Digest of the OSS-Fuzz base Docker image \\
error\_category            & Failure category per~\cite{nourry2025my} \\
root\_cause\_commit        & Commit identified as the root cause of the build failure \\
root\_cause\_workspace     & Workspace where the root cause is located \\
fixed\_state               & Whether the case is repaired by the evaluated system \\
\bottomrule
\end{tabular}%
}
\end{table}

\subsection{Failure Taxonomy and Distribution}

We classify the $128$ cases using the root-cause (RC) taxonomy of Nourry et al.~\cite{nourry2025my}, which defines $25$ root causes grouped into $11$ generic categories. Each category is specified by diagnostic criteria, representative build-log patterns, and boundary examples.

\textbf{Annotation Procedure.} Three authors annotated all $128$ cases following Nourry et al.~\cite{nourry2025my}. For each case, annotators inspected the build log, failure metadata, and failure-time repository state, and assigned one RC category. When the log was insufficient, they further examined the project repository, OSS-Fuzz issue tracker, related pull requests, and commit history. For logs with multiple failure patterns, we used the first fatal diagnostic for classification because later errors are often cascading effects.

Annotation proceeded through calibration, independent labeling, and adjudication. All annotators first labeled the same $30$ random cases to align category boundaries. The remaining $98$ cases were then double-labeled and disagreements were resolved by the full panel. Fleiss' $\kappa$ on the calibration set was $0.76$, indicating substantial agreement. All $128$ cases were classified, with none marked as insufficient evidence.

\textbf{Taxonomy Distribution.} Our dataset covers nine of the 25 RC categories in~\cite{nourry2025my}. For brevity, we denote these benchmark-local categories as RC1 to RC9 while preserving their definitions. Table~\ref{tab:rc_distribution} reports the distribution of the $128$ cases.

RC1 and RC2 dominate the dataset, accounting for $87.50\%$ of cases. Compilation and dependency labels cover both ordinary build problems and integration-specific constraints. The remaining $12.50\%$ of cases span seven smaller categories, each below $5\%$.

\textbf{Case Status.} All case-status statistics were collected before \mbox{June 15, 2026}. We divide the $128$ cases into two groups. \textit{SolvedCases} contains $60$ failures that had been repaired and successfully validated before this date. \textit{RealWorldCases} contains $68$ failures that remained active in the collected OSS-Fuzz records. This status split distinguishes known-fix and then-unresolved cases.

\textbf{Language Distribution.} The seven labels include the Java Virtual Machine (JVM) ecosystem. Their counts are C++ (45), JVM (21), C (19), Python (18), Rust (12), Go (11), and Swift (2), totaling 128 cases. C and C++ are separate labels. Multiple incidents can belong to one project, so cases are not necessarily statistically independent. Section~\ref{sec:category} reports repair rates by category.

\textbf{Benchmark Value and Scope.} Following Build-bench~\cite{buildbench}, the benchmark couples failure-time build contexts with executable verification. It targets continuous-fuzzing integration and covers nine of the 25 root-cause categories.

\section{Design of FuzzMedic}
\label{sec:approach}

\begin{figure*}[t]
\centering
\includegraphics[width=0.8\textwidth]{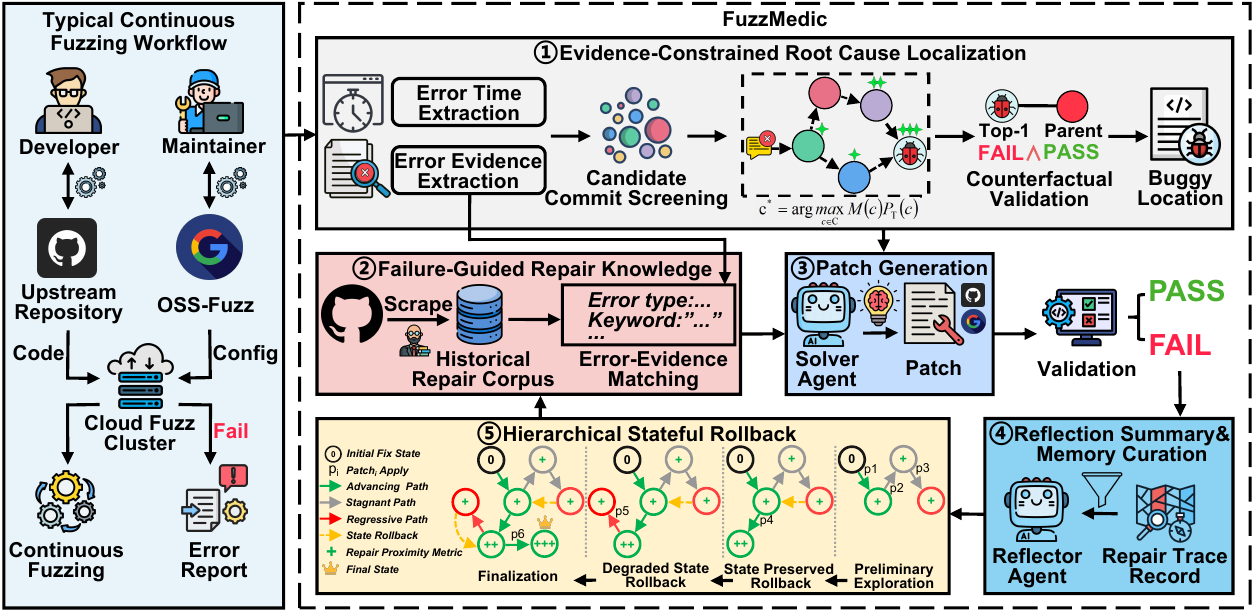}
\caption{Overview of the FuzzMedic architecture. Evidence localization, template-guided patch generation, Docker validation, trajectory memory, and state recovery form an iterative repair loop.}
\label{fig:overview}
\Description{The original FuzzMedic workflow links upstream and OSS-Fuzz workspaces to root-cause localization, historical knowledge, patch generation, validation, curated memory, and stateful rollback.}
\end{figure*}

FuzzMedic repairs OSS-Fuzz builds through collaborating agents (Figure~\ref{fig:overview}). ECRCL localizes candidate causes from logs and permitted repository history. FGRK supplies repair constraints before patch generation. Each applied patch is checked in the containerized build environment. RSMC retains validated outcomes as compact memory, while HSR tracks physical workspace states and recovers from unproductive branches. Diagnosis, knowledge, validation, memory, and state recovery thus form an iterative loop across upstream and OSS-Fuzz artifacts.

\subsection{Evidence-Constrained Root-Cause Localization}

Before patch generation, FuzzMedic must decide where the observed build failure should be repaired. This is non-trivial because the visible compiler, linker, or dependency error may be only a symptom, while the actionable fix may lie in another repository or configuration artifact. ECRCL addresses this problem by localizing the causal commit of a fuzzing build failure using only the build log and failure timestamp, without binary artifacts, coverage feedback, or runtime traces.

\textbf{Fault Evidence Abstraction.} ECRCL abstracts the build log into a \textit{primary fault evidence region}, which captures the earliest fatal diagnostic and filters cascading diagnostics.

\textbf{Candidate Commit Space Construction.} ECRCL builds an observation-constrained candidate set using spatial correlation, temporal proximity, and repair-scope determination. Spatial correlation retains commits that modify fault-relevant files, build configurations, dependency specifications, or structural code elements. Temporal proximity considers commits around the failure timestamp, reflecting that most OSS-Fuzz build failures are triggered by recent changes in upstream code, OSS-Fuzz configurations, or shared dependencies~\cite{nourry2025my}. Repair-scope determination estimates whether the failure originates from the upstream source or downstream fuzz-build configuration workspace.

ECRCL uses a two-tier localization strategy inspired by the change-inducing-fix analysis of Śliwerski, Zimmermann, and Zeller~\cite{sliwerski2005changes}. Precise Tracking extracts file paths and line numbers from diagnostics and applies \textit{git blame} to identify the responsible commit. Temporal Fallback handles logs without file-line evidence, including dependency-resolution and Docker failures. It retrieves commits from both repositories within one day before or after the failure and treats their modified files as suspects.

\textbf{Evidence Consistency Optimization.}
ECRCL refines candidates through PageRank-style bounded message passing over a directed heterogeneous evidence graph $G=(V,E,W)$. The node set $V$ contains the primary failure region, program and build entities, and candidate commits. Entities include files, lines, diff hunks, and dependencies. The edge set $E$ encodes directed evidence relations, while $W$ assigns each edge $(u,v)$ a non-negative strength $w(u,v)$. Relations connect failure regions to files and files or lines to commits. Let $C$ denote the candidate commit set and $N_{\text{region}}$ the primary failure-region node.

ECRCL initializes the evidence distribution from the primary failure region, with $P_0(N_{\text{region}})=1$ and $P_0(u)=0$ for all other nodes. For a node $u$, let $\mathcal{N}^{+}(u)$ and $\mathcal{N}^{-}(u)$ denote its outgoing and incoming neighbors, respectively. Edge weights are normalized into transition probabilities.
\begin{equation}
\Pr(u \rightarrow v)=
\frac{w(u,v)}
{\sum_{k\in \mathcal{N}^{+}(u)} w(u,k)} .
\end{equation}
ECRCL then performs bounded message passing for $T$ iterations.
\begin{equation}
P_t(u)=(1-d)I(u)+d\sum_{v\in \mathcal{N}^{-}(u)}
\Pr(v\rightarrow u)P_{t-1}(v),
\end{equation}
where $P_t(u)$ is the relevance mass of node $u$ after iteration $t$, $d\in(0,1)$ is the damping factor, and $I(u)$ is the restart vector with $I(N_{\text{region}})=1$ and $I(u)=0$ for all other nodes.

To suppress false positives from incompatible build environments, ECRCL applies a gating coefficient $M(c_i)\in[0,1]$ to each candidate commit $c_i\in C$. This coefficient summarizes compatibility with the current sanitizer, architecture, compiler, and toolchain configuration. The final score is
\begin{equation}
Score(c_i)=M(c_i)\,P_T(c_i),
\end{equation}
Candidate scores are sorted in descending order. The reported main experiment selects the top-ranked candidate ($k=1$) for counterfactual verification. An optional top-three verification mode is available but is not used for these reported results. No minimum-score or score-gap threshold is used. If no selected candidate passes verification, localization returns UNKNOWN.

\textbf{Fixed Settings.} Propagation uses $d=0.85$ and $T=3$. Edge weights are $1.0$ for message-to-commit links, $1.1$ for fault-region links to messages containing error/fail/fix, $1.3$ for matching file nodes, and $1.5$ for blame-anchor links to commits. Temporal file-to-commit links use $w_t$ below. Environment compatibility is binary, $M\in\{0,1\}$; the candidate cutoff is $k=1$, and HSR's neutral tolerance is $\theta=2$.

The temporal weight is
\begin{equation}
w_t=\max(0.1,1.5\,2^{-\Delta t/86400}),
\end{equation}
where $\Delta t$ is the absolute commit/failure time difference in seconds. Outgoing weights are normalized before propagation. Ranked candidates are hypotheses, not confirmed causes.

\textbf{Counterfactual Verification.}
ECRCL verifies ranked candidates through counterfactual build replay under the same OSS-Fuzz configuration. A commit is confirmed only if the parent revision builds, reapplying the candidate reproduces the observed build failure, and the reproduced failure matches the original primary fault evidence in both failure stage and diagnostic signature. The verified result is stored as a root-cause attribution artifact for patch generation.

\subsection{Failure-Guided Repair Knowledge}

After localization, the agent still needs repair knowledge that is often absent from raw logs. OSS-Fuzz builds impose constraints on container layout, sanitizer instrumentation, fuzzing engines, dependency setup, and fuzz-target packaging, so the same error message may require a fuzzing-specific fix rather than a generic build edit. FGRK addresses this gap by retrieving historically validated repair knowledge based on the current failure pattern and using it to guide patch generation.

\textbf{Project and Incident Separation.} FGRK sources cover December 29, 2022 to June 14, 2026, the same window as the benchmark. The cutoff applies to the actual failure and repair content used, including archived versions of evolving pages. Later additions and unverifiable versions are excluded. From 1,940 candidate source records, completeness/time checks and incident deduplication yield 502 historical failures from 210 normalized projects. Excluding all 253 records from the 86 source projects matching any of the 113 evaluation projects leaves 249 failures from 124 projects. Checks of forks, mirrors, shared upstreams, and patch propagation then remove 42 records from 14 related projects and seven additional same-origin incidents. A final provenance review excludes 12 records, leaving 188 failures from 110 projects (Appendix~\ref{app:fgrk}). Sharing a generic error type or build principle alone does not constitute incident overlap.

\textbf{Knowledge Construction.} The strict knowledge base is regenerated only from retained sources. Removing project names from mixed-source templates is insufficient. A large language model (LLM) assists extraction of failure patterns, applicability conditions, diagnoses, and general repair actions. Two authors independently review technical correctness, provenance, and case specificity. The result contains 43 error-pattern entries (ERR) and 13 reusable guideline entries (GL), totaling 56. Each entry retains its source mapping, but runtime templates omit evaluation incident identifiers, reference patches, and case-specific repair instructions. Reference fixes and source evidence remain in an audit environment inaccessible to the Solver agent, which generates candidate patches. This provides project-level separation with related-incident checks. It is not a temporal split preceding all evaluation cases.

\begin{table}[t]
\centering\caption{FGRK construction and source-selection ledger. Exclusions are sequential and non-overlapping. Initial materials become incident records after deduplication. The count 238 denotes preliminary project identifiers.}\label{tab:fgrk_flow}
\setlength{\tabcolsep}{3pt}
\begin{tabular}{p{.48\columnwidth}rrr}\toprule Stage & Excluded & Retained & Proj.\\\midrule
Initial failure/repair candidate collection &---&1,940&---\\
Time, completeness and version checks &620&1,320&238\\
Incident deduplication and project normalization &818&502&210\\
Exclude all records from evaluation projects &253&249&124\\
Related-repository and same-origin-incident audit &49&200&110\\
Final source-traceability review &12&188&110\\\midrule
Total excluded / final retained &1,752&188&110\\\bottomrule
\end{tabular}
\end{table}

\textbf{Reusable Patterns Across Projects.} The \textit{pffft}, \textit{varnish}, and \textit{netcdf} repairs correct the same Dockerfile pattern, where a COPY instruction with multiple sources lacks the destination's trailing slash. These cases illustrate how a general build constraint can recur across projects. FGRK captures such diagnostic conditions and repair principles, while its source-selection procedure separately excludes evaluation projects and related incidents. This distinguishes reusable build knowledge from transferring a case-specific reference patch.

A shared failure category can nevertheless require different repairs. The \textit{maven}, \textit{airflow}, and \textit{mdbook-i18n-helpers} cases are all dependency-related. The first requires a newer Maven version for plugin compatibility. The second requires a local development package version to be available during dependency resolution. The third updates a pinned nightly toolchain to handle a manifest requiring Rust Edition 2024. These examples connect the category-level labels to different actionable constraints. FGRK records applicability conditions and repair actions so guidance can be selected from the current diagnostic evidence rather than from an error-category label alone. Appendix~\ref{app:patches} links the \mbox{accepted repairs}.

\textbf{Retrieval and Use.} Retrieval first reads FAILURE\_REGION from \texttt{commit\_changed.txt}. When unavailable or empty, it uses a fixed tail slice of the current build log. Positive regular expressions must match and negative expressions must not match. Keywords in matching ERR entries associate relevant GL entries. Fixed priority, deduplication, and tie-breaking rules select at most four ERR/GL entries in total for the Solver prompt. Queries use only permitted failure-state and repair-attempt information. The knowledge base is frozen during evaluation. Repair memory and related caches are reset between cases and independent runs. Repository snapshots, mounts, and tool access exclude reference fixes and post-failure repair history. These controls do not rule out public repairs already present in closed-model pretraining.

\subsection{Reflection Summary and Memory Curation}

Because fuzzing-build repair requires repeated, expensive validation, later decisions must retain solved subproblems and failed strategies. Keeping complete logs and patch histories enlarges context and can obscure relevant evidence~\cite{liu2024lost}. RSMC converts each validated repair attempt into compact semantic memory.

Specifically, RSMC compares pre-patch and post-patch validation states to identify resolved symptoms, persistent problems, and priorities for the next round, producing a structured progress assessment inspired by verbal reinforcement~\cite{shinn2023reflexion}. It extracts verbose build logs into per-round summaries that preserve decision-relevant information while removing low-level noise. The curated context retains only extracted memory, reflection results, and patch-level artifacts, maintaining a persistent cross-round view of strategy evolution. This helps the agent avoid failed strategies and focus on the primary unresolved failure.

\subsection{Hierarchical Stateful Rollback}

Unlike ordinary patch generation, fuzzing build repair mutates a live workspace across rounds. An incorrect edit may introduce secondary failures, hide the original error, or cause later attempts to build on a polluted state. HSR addresses this non-monotonic feedback problem by constructing a repair-state graph (RSG), where each node records a validated workspace state and each edge captures the transition caused by a patch attempt.

\textbf{State Representation.} Each RSG node stores $S=\langle L,V\rangle$. The stage $L\in\{1,\dots,6\}$ represents six ordered pipeline phases, namely Environment Bootstrap, Dependency Resolution, Instrumented Compilation, Fuzzer Linkage, Artifact Validation, and Runtime Fuzz Execution. These correspond to L1 through L6. Advancement is a provisional search signal. A later failure may be newly introduced or newly exposed, and HSR cannot generally distinguish these causes.

The auxiliary score $V\in\{0,1,2,3\}$ counts passes among artifact recognition (\texttt{list\_fuzzers}), infrastructure compliance (\texttt{check\_build}), and runtime stability (\texttt{run\_fuzzer}). These checks guide state comparison. Only passing \texttt{check\_build} determines reported repair success, so the other checks impose no additional acceptance requirement.

Stage advancement is the primary progress signal, with validation changes providing auxiliary constraints. State dominance is defined as $S_a\succeq S_b$ if $L_a>L_b$, or if $L_a=L_b\land V_a\geq V_b$.

\begin{algorithm}[t]
\footnotesize
\caption{HSR rollback decision policy}
\label{alg:hsr_policy}
\KwIn{Repair-state graph $\mathcal{G}$, nodes $n_{prev}, n_{cur}$, neutral tolerance $\theta=2$}
\KwOut{Decision $a$, next-round initial state $n_{init}$}

$S_{prev} \leftarrow \langle L(n_{prev}), V(n_{prev}) \rangle$\;
$S_{cur}   \leftarrow \langle L(n_{cur}), V(n_{cur}) \rangle$\;
$(L_{prev}, V_{prev}) \leftarrow S_{prev}$\;
$(L_{cur}, V_{cur}) \leftarrow S_{cur}$\;

\BlankLine
\tcp{Stage progression or validation gain}
\If{$L_{cur} > L_{prev}$}{
  Accept($n_{cur}$)\;
  ResetNeutralCounter()\;
  \Return (Continue, $n_{cur}$)\;
}

\If{$L_{cur}=L_{prev} \;\land\; V_{cur}>V_{prev}$}{
  Accept($n_{cur}$)\;
  ResetNeutralCounter()\;
  \Return (Continue, $n_{cur}$)\;
}

\If{$L_{cur}=L_{prev} \;\land\; V_{cur}=V_{prev}$}{
  IncreaseNeutralCounter()\;
  \If{NeutralCounter() $< \theta$}{
    \Return (Observe, $n_{cur}$)\;
  }
}

\BlankLine
\tcp{Regression or repeated neutrality. Roll back to a dominating ancestor}
$n_{target} \leftarrow$ DominatingAncestor($\mathcal{G}, n_{cur}$)\;
InitNextRound($n_{target}$)\;
RelocalizeRootCause()\;
\Return (Rollback, $n_{target}$)\;
\end{algorithm}

\textbf{Rollback Decision.} Algorithm~\ref{alg:hsr_policy} formalizes the HSR rollback policy, which compares $S_{cur}$ with $S_{prev}$ after each round to select the starting state for the next iteration. Four outcomes are distinguished. First, stage progression ($L_{cur} > L_{prev}$) starts from $n_{cur}$. Second, a stable stage with improved validation ($L_{cur} = L_{prev} \land V_{cur} > V_{prev}$) starts from $n_{cur}$. Third, a stable stage with no gain ($L_{cur} = L_{prev} \land V_{cur} = V_{prev}$) is observed for one round and triggers rollback on the second consecutive neutral round ($\theta=2$). Fourth, stage regression, or validation decline at the same stage, rolls back to a dominating ancestor. This stage-first rule matches Algorithm~\ref{alg:hsr_policy}. A validation decrease accompanying a stage increase does not override the stage-first heuristic.

\textbf{Rollback Target Selection.} HSR searches upward for the nearest ancestor that dominates the current state under $\succeq$. If no suitable non-root ancestor exists, it returns to the root workspace. Both repositories are physically restored to the target node's recorded commit identifiers. The fixed neutral tolerance $\theta=2$ allows one unchanged round before rollback.

\textbf{Budget Controls.} There is no independent per-node attempt limit. Search is bounded by global retries, internal repair rounds, and project-level timeout. The reported main configuration permits eight rounds and 1.5 hours per case. The exact global retry configuration belongs in the artifact, rather than being inferred as a node-level limit.

\subsection{Implementation}

FuzzMedic is implemented in Python using Google ADK. A sequential preparation workflow initializes the historical environment, validates the failure, localizes candidate causes, and prepares repair context. A loop then coordinates patch generation, application, validation, memory updates, and rollback. Agents communicate through structured prompts and a persistent repair trace. Workspace snapshots associate upstream and OSS-Fuzz revisions with validation outcomes so rollback restores physical repository state as well as semantic context. It uses DeepSeek-V4-Flash~\cite{deepseek_v4} as the foundation model for all agents, with the reported seed setting $42$. Agents for environment initialization, build execution, root-cause localization, rollback decision, and patch application use near-deterministic decoding settings ($\text{temperature}=0.0$, $\text{top\_p}=0.1$) to maximize consistency. The RSMC and prompt assembly agents use slightly higher settings ($\text{temperature}=0.2$, $\text{top\_p}=0.3$) to support reflective reasoning and context synthesis.

The loop stops when \texttt{check\_build} succeeds, eight repair rounds are exhausted, or the 1.5-hour case budget expires. Passing \texttt{check\_build} is the sole reported success criterion. The other checks are reference signals for search and HSR state scoring, not additional acceptance gates. Search has no proof that stage advancement is monotonic correctness progress. Complete validation remains necessary.

\section{Evaluation}
We address four research questions (RQs) on repair effectiveness (RQ1), resource requirements (RQ2), component ablations (RQ3), and model/budget sensitivity (RQ4). Main benchmark results are distinguished from supplementary pilots. Independent repetitions quantify observed run variation.

\subsection{Setup and Measurement}
\textbf{Benchmark and Baselines.} All 128 cases are counted in the repair-rate denominator. The native-budget comparison uses a shared Docker execution backend and \texttt{check\_build} criterion. Agentless~\cite{xia2024agentless} scans both upstream and OSS-Fuzz trees and generates six patch candidates per case. CompileAgent~\cite{hu2025compileagent} is adapted to fuzz-build validation with a 40-step budget. SWE-Search~\cite{antoniades2024swe} searches both repositories through a \textit{FuzzBuild} action with a 20-round Monte Carlo tree search (MCTS) budget. In the reported model configuration, FuzzMedic, Agentless, and CompileAgent use DeepSeek-V4-Flash. SWE-Search uses DeepSeek-V4-Flash, GPT-5.5 Instant, and Claude Sonnet 4.6 for trajectory discrimination. These native budgets are not equivalent in token, time, interaction, or build-attempt consumption.

\textbf{Success and Costs.} A repair succeeds when its patch passes OSS-Fuzz's \texttt{check\_build} in the reconstructed context. Artifact recognition and runtime stability contribute to HSR's auxiliary score $V$. Token counts and elapsed times are medians over successfully repaired cases, while model-service cost is the mean over those cases.

\textbf{Configuration Selection.} We scanned 42 round/time-budget configurations on the 128-case benchmark and selected eight rounds and 1.5 hours per case. The parameter scan and main comparison were executed separately.

\textbf{Independent Repetitions.} Each main method and ablation configuration was run independently three times on the fixed evaluation set. For FuzzMedic, the mean repair rate is 48.44\%, the sample standard deviation is 1.56 percentage points, and the coefficient of variation is 3.23\%. The 95\% confidence interval for the mean is [44.56\%, 52.32\%]. It is calculated from the three run-level rates using Student's $t$ distribution with two degrees of freedom.

\textbf{Knowledge-Base Version.} The main experiment and applicable ablations use the frozen strict FGRK version specified in Section~\ref{sec:approach}. It contains 56 entries and excludes evaluation projects and related incidents. The FGRK ablation omits these entries.

\subsection{Repair Effectiveness (RQ1)}
\begin{figure}[t]
\centering
\includegraphics[width=0.8\columnwidth]{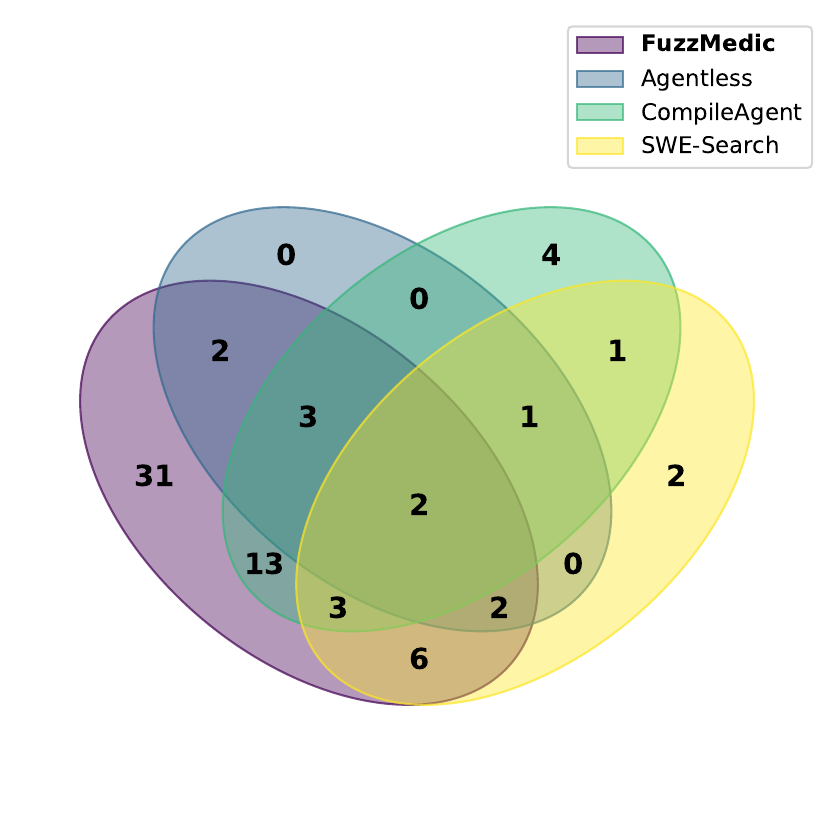}
\caption{Overlap of successfully repaired cases among FuzzMedic and the three baselines. Each number counts cases in one exclusive region.}
\label{fig:main_baselines_venn}
\end{figure}

\begin{figure}[t]
\centering
\includegraphics[width=\columnwidth]{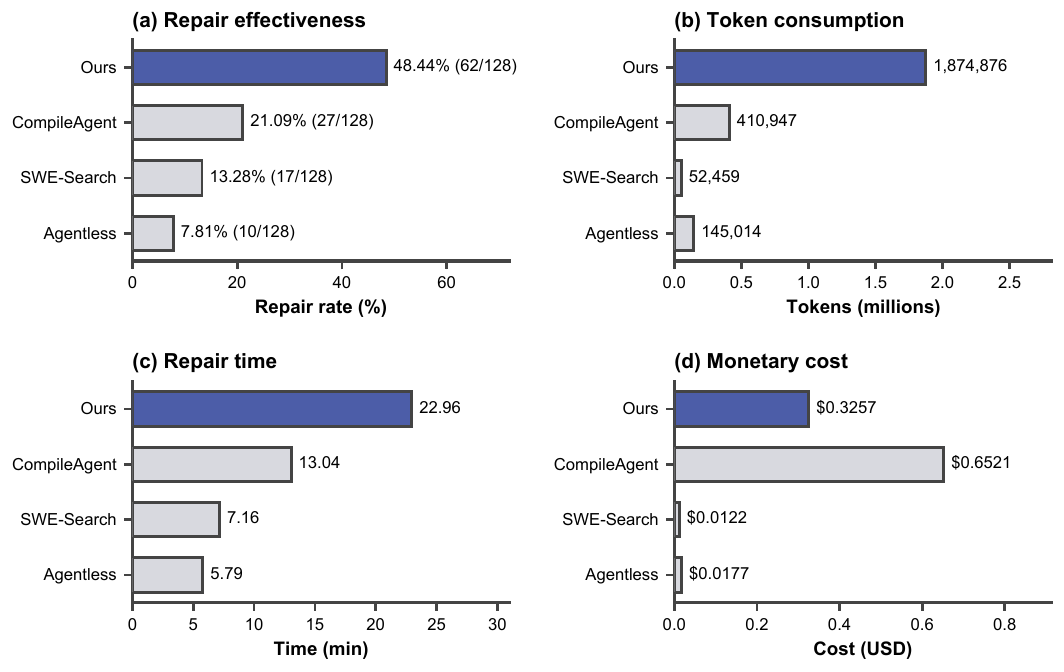}
\caption{Repair effectiveness and resource consumption on the 128-case benchmark. Token counts and elapsed times are medians over successful cases; monetary costs are means over successful cases.}
\label{fig:main_results}
\end{figure}

Figure~\ref{fig:main_results} reports 62/128 restored builds (48.44\%) for FuzzMedic, 27/128 (21.09\%) for CompileAgent, 17/128 (13.28\%) for SWE-Search, and 10/128 (7.81\%) for Agentless. The gap to the strongest native-budget baseline is 27.35 percentage points after rounding. Figure~\ref{fig:main_baselines_venn} shows the repair overlap, including 31 cases repaired by FuzzMedic and missed by all three baselines.
\subsubsection{Outcomes by Failure Category}\label{sec:category}
\begin{figure}[t]
\centering
\includegraphics[width=\columnwidth]{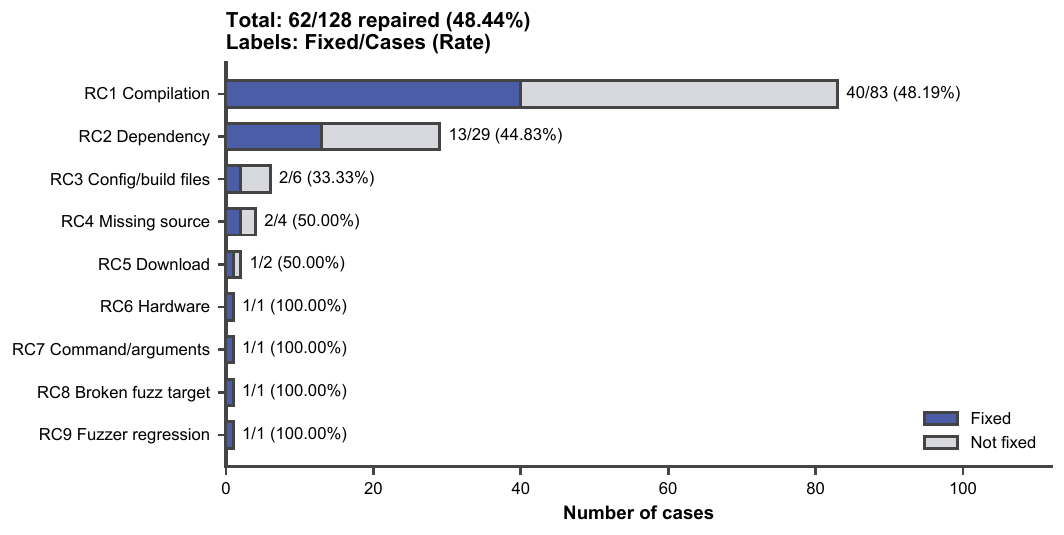}
\caption{Repair outcomes by benchmark-local root cause. Labels show repaired cases, category totals, and repair rates. Small-category rates are descriptive.}
\label{fig:category_results}
\end{figure}

Figure~\ref{fig:category_results} partitions the 128 cases and 62 repairs by category. Each category's repair rate uses its own case count as the denominator. Compilation and dependency failures account for 112 cases (87.50\%), so they dominate the aggregate result. RC5--RC9 contain at most two cases each. Their rates do not support reliable category-level generalization or show that rare failures are easier.

\subsubsection{Matched Caps and Additional Comparisons}
Table~\ref{tab:matched} compares the systems using a unified model and matched per-case token/time caps, where M denotes one million tokens. CompileAgent repairs more cases at the lowest cap, while FuzzMedic repairs more at higher caps. Under the 8M-token/90-minute cap, FuzzMedic repairs 59 cases. Build and interaction budgets are \mbox{not matched}.
\begin{table}[t]
\centering\caption{Matched token/time-cap pilot on 128 cases, with unmatched build and interaction budgets.}\label{tab:matched}
\begin{tabular}{lrr}\toprule Per-case token/time cap & FuzzMedic & CompileAgent\\\midrule
1M / 15 minutes &11 (8.59\%) &16 (12.50\%)\\
4M / 60 minutes &50 (39.06\%) &22 (17.19\%)\\
8M / 90 minutes &59 (46.09\%) &25 (19.53\%)\\\bottomrule
\end{tabular}
\end{table}

Additional experiments compare FuzzMedic with a bare iterative agent using the same model, validator, and round budget on 64 cases. The bare agent repairs four cases, compared with 35 for FuzzMedic. On 20 applicable C/C++ compilation failures, CXXCrafter repairs five cases, compared with 12 for FuzzMedic.

\textbf{Failure Analysis.} Among failed attempts entering the repair loop, 41.8\% record root-cause localization and 58.2\% lack it. Localization can still leave coupled dependency and integration constraints unresolved. In \textit{quiche}, edits to an outdated patch introduce malformed hunks before exposing Bazel/FuzzTest incompatibilities. Recurring toolchain and path errors further prevent convergence. These traces reveal failures in translating diagnosis into compatible edits, constructing patches, and grounding repair in clear evidence.

\begin{center}
\colorbox{gray!20}{%
\parbox{\dimexpr0.97\linewidth-2\fboxsep\relax}{%
\textbf{Answer to RQ1.} FuzzMedic repairs 62 of 128 cases (48.44\%), including 31 cases missed by all three baselines. Its repair rate exceeds that of the strongest evaluated baseline, CompileAgent, by 27.35 percentage points.
}}
\end{center}

\subsection{Resource Requirements (RQ2)}
Figure~\ref{fig:main_results} also compares resource requirements. Among successfully repaired cases, FuzzMedic uses a median of 1,874,876 tokens and 22.96 minutes, with a mean model-service cost of \$0.3257. CompileAgent uses fewer tokens and less time on its successful cases, whereas FuzzMedic achieves a higher repair rate.

The median successful patch modifies one file and five lines. Root-cause localization succeeds in 38.71\% of repaired cases, while 50.00\% involve rollback.

\begin{center}
\colorbox{gray!20}{%
\parbox{\dimexpr0.97\linewidth-2\fboxsep\relax}{%
\textbf{Answer to RQ2.} Successful repairs require a median of 1.87 million tokens and 22.96 minutes, with a mean model-service cost of \$0.3257. The median patch changes one file and five lines.
}}
\end{center}

\subsection{Component Ablations (RQ3)}
Table~\ref{tab:ablation} shows that every component removal reduces repairs and raises successful-case median token use by approximately 1.24--1.48 times. HSR has the largest effect, reducing repairs from 62 to 42 when removed. Its recovery of earlier workspace states complements RSMC's retention of repair history, allowing subsequent edits to preserve progress while avoiding repeated unsuccessful strategies.

\begin{figure}[t]
\centering
\includegraphics[width=0.8\columnwidth]{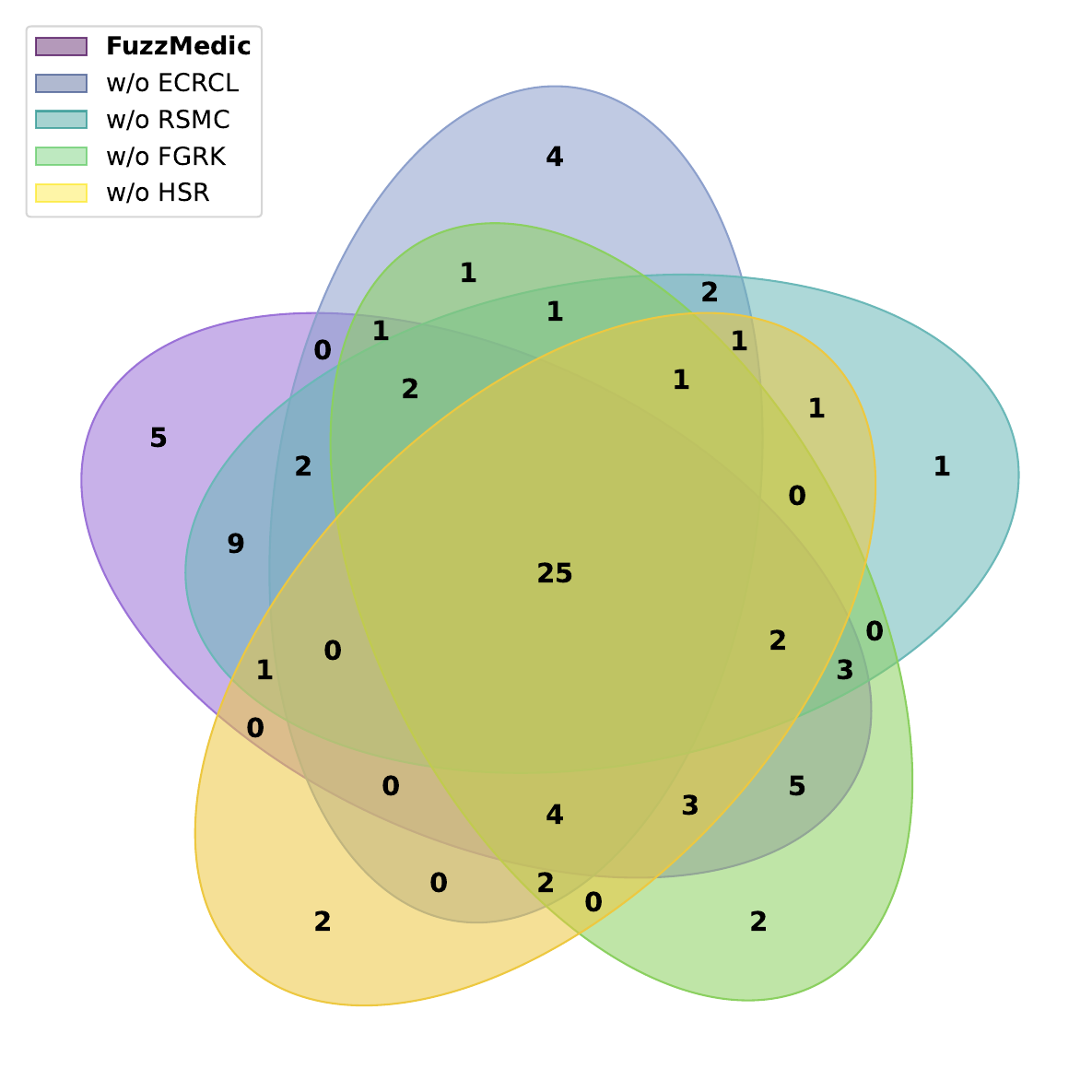}
\caption{Overlap of repaired cases between the full system and the four ablations. Each number counts cases in one exclusive region. The legend's Few-shot RAG denotes FGRK.}
\label{fig:main_ablations_venn}
\end{figure}

ECRCL and FGRK complement this recovery through evidence-guided localization and reusable build constraints before patch generation. Their removal also lowers repair counts, supporting the value of both diagnostic evidence and domain guidance. Figure~\ref{fig:main_ablations_venn} shows 25 cases repaired by all configurations and five repaired only by the full system, illustrating differences in the cases reached by each configuration.

\begin{table}[t]
\centering
\caption{Component ablations on the 128-case benchmark. Tokens and time are successful-case medians.}
\label{tab:ablation}
\setlength{\tabcolsep}{4pt}
\resizebox{\columnwidth}{!}{%
\begin{tabular}{lcccc}
\toprule
\textbf{Variant} & \textbf{Fixed} & \textbf{Rate (\%)} & \textbf{Tokens} & \textbf{Time (min)} \\
\midrule
FuzzMedic (Full)    & \textbf{62} & \textbf{48.44} & \textbf{1,874,876} & 22.96 \\
\midrule
w/o RSMC            & 51 & 39.84 & 2,448,098 & 24.13 \\
w/o HSR             & 42 & 32.81 & 2,777,840 & 27.41 \\
w/o ECRCL           & 46 & 35.94 & 2,328,114 & 23.63 \\
w/o FGRK            & 52 & 40.62 & 2,444,078 & 24.67 \\
\bottomrule
\end{tabular}%
}
\end{table}
\begin{center}
\colorbox{gray!20}{%
\parbox{\dimexpr0.97\linewidth-2\fboxsep\relax}{%
\textbf{Answer to RQ3.} Removing any component reduces repair effectiveness and increases median token use per successful repair. HSR has the largest observed effect, with its removal reducing repairs from 62 to 42. The other removals yield 46--52 repairs.
}}
\end{center}

\subsection{Model and Budget Sensitivity (RQ4)}
Table~\ref{tab:model_comparison} compares DeepSeek-V4-Flash, Claude Sonnet 4.6~\cite{anthropic_claude}, and GPT-5.5 Instant~\cite{openai_gpt55} using identical prompts, tools, and budgets. Repair rates span 4.69 percentage points. Claude Sonnet 4.6 achieves the highest repair rate and the lowest successful-case median token use and elapsed time.
\begin{table}[t]
\centering
\caption{Foundation-model comparison. Tokens and time are successful-case medians.}
\label{tab:model_comparison}
\setlength{\tabcolsep}{4pt}
\resizebox{\columnwidth}{!}{%
\begin{tabular}{lcccc}
\toprule
\textbf{Foundation model} & \textbf{Fixed} & \textbf{Rate (\%)} & \textbf{Tokens} & \textbf{Time (min)} \\
\midrule
DeepSeek-V4-Flash & 62 & 48.44 & 1,874,876 & 22.96 \\
Claude Sonnet 4.6  & 64 & 50.00 & 1,122,970 & 20.07 \\
GPT-5.5 Instant    & 58 & 45.31 & 1,668,605 & 22.71 \\
\bottomrule
\end{tabular}%
}
\end{table}
We also evaluate a locally deployed open-weight DeepSeek-V4-Flash-0731 model on 32 cases. It repairs 17 cases (53.12\%) using eight A100 graphics processing units (GPUs), each with 80 GB of memory, connected through NVLink. This experiment demonstrates that FuzzMedic can perform build repair with a locally hosted model.

\begin{figure}[t]
\centering
\includegraphics[width=\columnwidth]{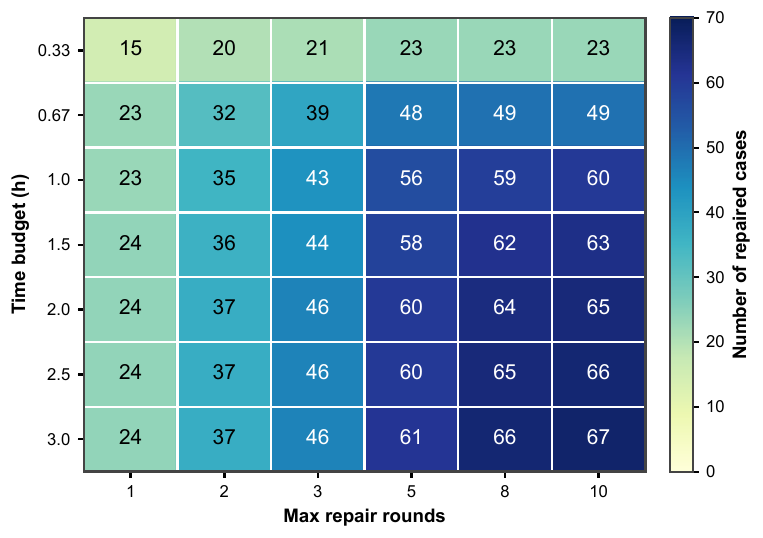}
\caption{Repaired projects under varying round and time budgets. Counts refer to projects rather than failure cases.}
\label{fig:heatmap}
\Description{Heatmap of repaired-project counts for six round limits and seven time budgets. The reported default is eight rounds and 1.5 hours.}
\end{figure}

Figure~\ref{fig:heatmap} shows project-level repair outcomes across 42 combinations of round limits and time budgets. With eight rounds, increasing the time budget from 1.0 to 1.5 and 2.0 hours raises the number of repaired projects from 59 to 62 and 64, respectively. The highest count is 67 projects with ten rounds and three hours. We select eight rounds and 1.5 hours as the default to balance repair effectiveness and execution time. Increasing the time budget from 1.5 to three hours at eight rounds yields three additional repaired projects, indicating diminishing returns from longer execution.
\begin{center}
\colorbox{gray!20}{%
\parbox{\dimexpr0.97\linewidth-2\fboxsep\relax}{%
\textbf{Answer to RQ4.} Repair rates range from 45.31\% to 50.00\% across the three evaluated models. The budget sweep shows diminishing returns from longer execution, supporting eight rounds and 1.5 hours as a practical effectiveness--time trade-off on this benchmark.
}}
\end{center}


\section{Real-World Impact}\label{sec:impact}
\subsection{PR-Experiment Workflow}
The workflow was submitted as a PR to Google's \textit{oss-fuzz-gen} repository and exercised through its PR-experiment mechanism. It covers build-log acquisition, failure-metadata extraction, automated repair with build validation, and production of PR information. Human verification precedes external patch submission. Maintainers review and merge the resulting changes. This experiment does not establish adoption as an unattended production service or inclusion in \textit{oss-fuzz-gen}'s main branch.

\subsection{Maintainer-Merged Repairs}
The case register contains 29 merged patches across 28 projects. Of these, 10 patches (34.48\%) repair upstream defects and 19 (65.52\%) repair OSS-Fuzz integration issues, demonstrating contributions to both project code and fuzzing infrastructure. Repairs are dated November 12, 2025 through September 20, 2026. Appendix~\ref{app:patches} lists the project, dates, failure category, repair location, and supporting links for each patch.

\begin{figure}[t]
\centering
\includegraphics[width=\columnwidth]{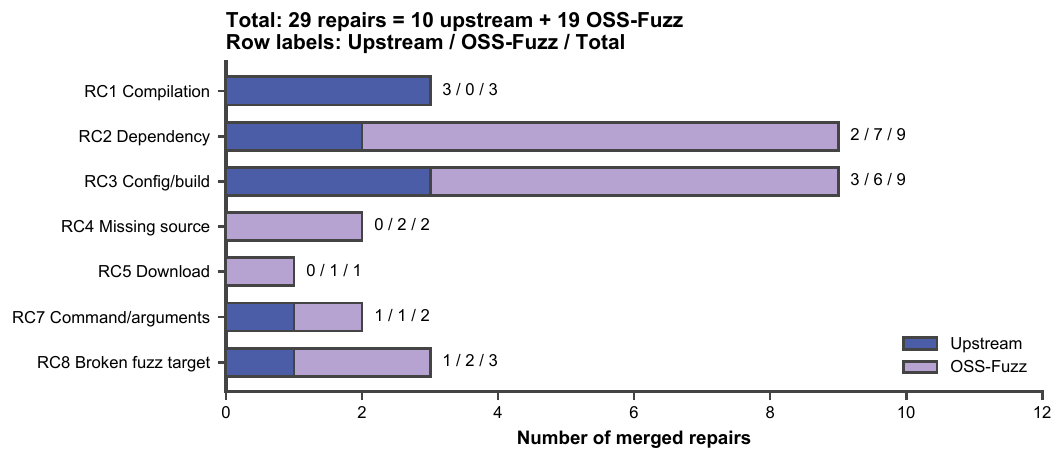}
\caption{The 29 maintainer-merged repairs grouped by root cause and repair location. Row labels give upstream, OSS-Fuzz, and total counts.}
\label{fig:merged}
\end{figure}

Figure~\ref{fig:merged} summarizes the 29 merged repairs, including three long-standing failures that had remained unresolved for more than one year and two cases that had never built successfully. Source or upstream build defects are reviewed in the upstream repository. Integration changes to Dockerfiles, scripts, or harnesses are reviewed through OSS-Fuzz. OSS-Fuzz-side acceptance is evidence of integration-maintenance utility, while upstream acceptance establishes adoption in the upstream project. The three long-standing failures affect \textit{airflow}, \textit{toml\_edit}, and \textit{mdbook-i18n-helpers}, with failure-to-repair intervals of 510, 449, and 470 days, respectively. The two previously unbuilt cases are \textit{bc-gh} and \textit{bazel-rules-fuzzing-test}. These repairs restore the build prerequisite for continuous fuzzing in projects affected by persistent or previously unresolved integration failures.

\section{Discussion and Threats to Validity}
\textbf{Benchmark Scope.} Retaining reproducible failures favors recoverable historical environments and accessible version metadata. The same 128-case benchmark was used for parameter selection and evaluation, which may yield optimistic estimates. Compilation and dependency failures dominate the benchmark, and several other categories contain only one or two cases.

\textbf{Knowledge Separation.} The frozen FGRK excludes evaluation projects and related incidents. Runtime access excludes reference patches, and memory is reset between cases and runs. These controls reduce direct leakage, but cannot completely exclude public repairs from closed-model pretraining.

\textbf{Variation and Comparisons.} Three complete runs characterize observed run-level variation, with the available uncertainty summary covering the full system. Native search budgets differ across methods, while matched token/time caps leave build and interaction counts unmatched. Resource summaries condition on successful cases and exclude unsuccessful-attempt expenditure.

\textbf{Repair Criterion.} Passing \texttt{check\_build} establishes build restoration, rather than preservation of harness coverage or semantic equivalence. HSR's stage ordering guides search, but later-stage failures can be newly introduced or newly exposed.

\section{Related Work}
\textbf{Fuzzing and Build Failures.} OSS-Fuzz automates sanitizer-instrumented fuzzing~\cite{ossfuzz,serebryany2017oss}. Empirical studies characterize OSS-Fuzz failures~\cite{nourry2025my}, build errors at Google~\cite{seo2014programmers}, compiler errors~\cite{zhang2019compiler}, Travis CI failures~\cite{beller2017oops}, and CI adoption~\cite{hilton2016usage}. Sharma et al. investigate efficient CI/CD fuzzing~\cite{sharma2024effective}. Build-bench studies package building across architectures with executable verification~\cite{buildbench}.

\textbf{Build Repair.} BuildMedic repairs Maven dependencies~\cite{macho2018automatically}, HireBuild uses build history~\cite{hassan2018hirebuild}, and CXXCrafter and CompileAgent automate software building~\cite{yu2025cxxcrafter,hu2025compileagent}. Agentless and SWE-Search provide repository-repair localization and search~\cite{xia2024agentless,antoniades2024swe}. FuzzMedic instead coordinates upstream and OSS-Fuzz artifacts with official build validation.

\textbf{Program and Vulnerability Repair.} GenProg and GenProgJS use test-based repair~\cite{le2011genprog,csuvik2025genprogjs}. KNOD incorporates domain knowledge~\cite{jiang2023knod}, Tare uses typing constraints~\cite{zhu2023tare}, ITER repairs multiple locations iteratively~\cite{ye2024iter}, and hybrid repair combines LLM-generated patches with program analysis~\cite{li2025hybrid}. Logs In, Patches Out uses sanitizer logs~\cite{kim2025logs}, APPATCH uses adaptive prompting~\cite{nong2025appatch}, and PATCHAGENT integrates localization and patch verification~\cite{yu2025patchagent}. FuzzMedic repairs the earlier build stage, before a runnable fuzzer and crash input are available.

\textbf{Repair Agents and Evaluation.} AutoCodeRover, RepairAgent, and HAFixAgent explore autonomous or history-aware repair~\cite{zhang2024autocoderover,bouzenia2025repairagent,shi2025hafixagent}. SWE-bench evaluates repository-issue resolution~\cite{jimenez2024swebench}, while OSS-Fuzz vulnerability repair starts from security findings~\cite{zhang2024fixing}. Passerine distinguishes test-passing patches from semantic correctness~\cite{passerine}.

\textbf{Agent Feedback and Analysis.} ReAct combines reasoning and actions~\cite{yao2022react}, Reflexion uses verbal feedback~\cite{shinn2023reflexion}, and long-context research motivates retaining relevant evidence~\cite{liu2024lost}. FuzzMedic applies these ideas through build-state memory and physical rollback. AutoEmpirical automates empirical fault analysis~\cite{autoempirical}.

\section{Conclusion}
FuzzMedic repairs OSS-Fuzz builds using localization, knowledge, memory, and rollback. It repairs 62 of 128 cases and yields 29 merged patches across 28 projects. Ablations support all four components, especially HSR. Failed attempts expose coupled constraints and ineffective edits. The PR experiment links log acquisition, validated repair, and PR-information generation.
\section{Data Availability}
Appendices~\ref{app:patches} and~\ref{app:selection} provide the 29 merged-patch records and the benchmark and knowledge-base construction audit, respectively.
\label{end:main}
\clearpage



\clearpage
\appendix
\setkeys{acmart.cls}{balance=false}
\nobalance
\onecolumn
\section{Merged-Patch Register}\label{app:patches}
Table~\ref{tab:patch_register} lists the 29 merged patches across 28 projects, comprising 10 upstream fixes and 19 OSS-Fuzz integration fixes. Types use the benchmark-local RC1--RC9 definitions in Table~\ref{tab:rc_distribution}. Dates are source-register dates, not agent runtimes. ``None yet'' indicates that the project has never built successfully. PR/issue links document review context, while the final column links to the fix/merge evidence. The \textit{glaze} record supplies a PR event rather than a canonical commit, marked explicitly.

\begingroup
\small
\setlength{\tabcolsep}{3pt}
\renewcommand{\arraystretch}{1.25}
\begin{longtable}{@{}r >{\raggedright\arraybackslash}p{90pt} ll c >{\raggedright\arraybackslash}p{90pt} >{\raggedright\arraybackslash}p{65pt}@{}}
\caption{All 29 maintainer-merged repairs, using this paper's RC1--RC9 category mapping.}\label{tab:patch_register}\\
\toprule
Case & Project & Failure date & Repair date & Type & Repair location & Fix/merge evidence\\
\midrule
\endfirsthead
\multicolumn{7}{l}{\textit{Table~\ref{tab:patch_register} continued}}\\
\toprule
Case & Project & Failure date & Repair date & Type & Repair location & Fix/merge evidence\\
\midrule
\endhead
\midrule\multicolumn{7}{r}{\textit{Continued on next page}}\\
\endfoot
\bottomrule
\endlastfoot
1 & \textit{sigstore} & 2025-11-05 & 2025-11-12 & RC1 & Upstream & \href{https://github.com/sigstore/sigstore/commit/d8ab8afb132621efe295937f5178b2d02938bf20}{\texttt{d8ab8afb}}\\
2 & \textit{mupdf} & 2025-11-08 & 2025-11-19 & RC5 & OSS-Fuzz (Dockerfile) & \href{https://github.com/google/oss-fuzz/commit/5ac272f888ce9e0b8692ec3057356a4ca4de793d}{\texttt{5ac272f8}}\\
3 & \textit{neqo} & 2025-11-03 & 2025-11-22 & RC4 & OSS-Fuzz (Dockerfile) & \href{https://github.com/google/oss-fuzz/commit/261e9e8784d317c6697a57e49d0bd914295a32b6}{\texttt{261e9e87}}\\
4 & \textit{croaring} & 2025-10-29 & 2025-11-21 & RC3 & Upstream & \href{https://github.com/RoaringBitmap/CRoaring/commit/d5f843304265e41b5d4ae3104f636f28f858c9d8}{\texttt{d5f84330}}\\
5 & \textit{maven} & 2025-11-25 & 2025-12-02 & RC2 & OSS-Fuzz (Dockerfile) & \href{https://github.com/google/oss-fuzz/commit/edbab81b93be32075b2af7c45041b48cc17c3234}{\texttt{edbab81b}}\\
6 & \textit{dgraph} & 2025-12-04 & 2025-12-12 & RC3 & OSS-Fuzz (build.sh) & \href{https://github.com/google/oss-fuzz/commit/2ddf9dba55086dca0c392e583b29d60f3e6d2247}{\texttt{2ddf9dba}}\\
7 & \textit{airflow} & 2024-07-20 & 2025-12-12 & RC2 & OSS-Fuzz (Dockerfile) & \href{https://github.com/google/oss-fuzz/commit/966a66250cc71a50fcc4577d5efb6e4ecc605eba}{\texttt{966a6625}}\\
8 & \textit{xstream} & 2025-11-07 & 2025-12-12 & RC3 & OSS-Fuzz (build.sh) & \href{https://github.com/google/oss-fuzz/commit/88c5240080d7d0a0feede9cfb8290e10741e7c64}{\texttt{88c52400}}\\
9 & \textit{pacemaker} & 2025-11-27 & 2025-12-13 & RC2 & OSS-Fuzz (build.sh) & \href{https://github.com/google/oss-fuzz/commit/a07a94be68415f4b9d99c1e9ce11af50ed809fad}{\texttt{a07a94be}}\\
10 & \textit{flyway} & 2025-12-09 & 2025-12-14 & RC2 & OSS-Fuzz (build.sh) & \href{https://github.com/google/oss-fuzz/commit/ff36f234e008e131adfd4f7444d47d8c911c6ef3}{\texttt{ff36f234}}\\
11 & \textit{mysql-server} & 2025-12-04 & 2025-12-18 & RC2 & OSS-Fuzz (Dockerfile) & \href{https://github.com/google/oss-fuzz/commit/134af986af46f5ab201feb4076a2e1f51758795e}{\texttt{134af986}}\\
12 & \textit{mosquitto} & 2025-10-01 & 2025-12-18 & RC3 & OSS-Fuzz (Dockerfile) & \href{https://github.com/google/oss-fuzz/commit/783f398540c701581c0d6a5e21ff7afb4ba61d73}{\texttt{783f3985}}\\
13 & \textit{airflow} & 2025-12-18 & 2025-12-23 & RC8 & OSS-Fuzz (build.sh) & \href{https://github.com/google/oss-fuzz/commit/f4f5e12b8a772d8ee39020a888b194cbd9d103ee}{\texttt{f4f5e12b}}\\
14 & \textit{pffft} & 2026-01-05 & 2026-01-14 & RC3 & OSS-Fuzz (Dockerfile) & \href{https://github.com/google/oss-fuzz/commit/722491e7bad3b63f61a302a8a04dfa1309875c9f}{\texttt{722491e7}}\\
15 & \textit{varnish} & 2026-01-05 & 2026-01-14 & RC3 & OSS-Fuzz (Dockerfile) & \href{https://github.com/google/oss-fuzz/commit/24a5c105caf5ad267aca7acb02bce9040cc4ec15}{\texttt{24a5c105}}\\
16 & \textit{netcdf} & 2026-01-05 & 2026-01-14 & RC3 & OSS-Fuzz (Dockerfile) & \href{https://github.com/google/oss-fuzz/commit/be6d6d7bb385059667b4b20a31d29b0cfcd2c1f5}{\texttt{be6d6d7b}}\\
17 & \textit{aptos-core} & 2026-01-30 & 2026-05-07 & RC1 & Upstream & \href{https://github.com/aptos-labs/aptos-core/commit/44a6cf440654cef3f124789e88eab7f37f6e8f93}{\texttt{44a6cf44}}\\
18 & \textit{roaring-bitmap} & 2025-08-22 & 2026-07-20 & RC3 & Upstream & \href{https://github.com/RoaringBitmap/RoaringBitmap/commit/e140aee9358b1e33be6e5b0cd8f043daef05299a}{\texttt{e140aee9}}\\
19 & \textit{sigstore-java} & 2025-11-11 & 2026-07-21 & RC2 & OSS-Fuzz (build.sh) & \href{https://github.com/google/oss-fuzz/commit/d944b51503f8b6cd5a43534a66f616fce41eb5e0}{\texttt{d944b515}}\\
20 & \textit{jimfs} & 2025-10-28 & 2026-07-21 & RC2 & Upstream & \href{https://github.com/google/jimfs/commit/7b8bcf810dc430a0abcbd84fa92975189b037306}{\texttt{7b8bcf81}}\\
21 & \textit{toml\_edit} & 2025-05-01 & 2026-07-24 & RC7 & OSS-Fuzz (build.sh) & \href{https://github.com/google/oss-fuzz/commit/c3deaec3e671bb2289cd1366af9dfc15c131f86c}{\texttt{c3deaec3}}\\
22 & \textit{glaze} & 2025-12-19 & 2026-08-03 & RC1 & Upstream & \href{https://github.com/stephenberry/glaze/pull/2752#event-28859172870}{PR event}\\
23 & \textit{bc-gh} & None yet & 2026-08-08 & RC7 & Upstream & \href{https://github.com/gavinhoward/bc/commit/d16fbee9f7b6367f6db0e46a9fd7dd3fd6f65b72}{\texttt{d16fbee9}}\\
24 & \textit{uint256} & 2026-03-01 & 2026-08-24 & RC8 & Upstream & \href{https://github.com/holiman/uint256/commit/3b6e9cdebb7a0462a2748bf24d7fac2b8a943666}{\texttt{3b6e9cde}}\\
25 & \textit{tomcat} & 2026-08-12 & 2026-09-15 & RC8 & OSS-Fuzz (code) & \href{https://github.com/google/oss-fuzz/commit/5aed4560637a86f22ab6514c6fa451fea5019310}{\texttt{5aed4560}}\\
26 & \textit{compress} & 2026-09-04 & 2026-09-16 & RC3 & Upstream & \href{https://github.com/klauspost/compress/commit/de8f55df9fb3af7aec787942d6b3c71e045bfee9}{\texttt{de8f55df}}\\
27 & \textit{mdbook-i18n-helpers} & 2025-06-03 & 2026-09-16 & RC2 & OSS-Fuzz (Dockerfile) & \href{https://github.com/google/oss-fuzz/commit/1d2405cfbf4e7819cb5d5a73aa8dfc5b55894e35}{\texttt{1d2405cf}}\\
28 & \textit{libunwind} & 2026-08-29 & 2026-09-16 & RC4 & OSS-Fuzz (build.sh) & \href{https://github.com/google/oss-fuzz/commit/70a468cd8a80cfae2f4db3316cb2476835d6618e}{\texttt{70a468cd}}\\
29 & \textit{bazel-rules-fuzzing-test} & None yet & 2026-09-20 & RC2 & Upstream & \href{https://github.com/bazel-contrib/rules_fuzzing/commit/faf15ee0b867f10222015674dac271230b57546e}{\texttt{faf15ee0}}\\
\end{longtable}
\endgroup
\twocolumn

\section{Benchmark and Knowledge-Base Construction Audit}\label{app:selection}
\subsection{Case Metadata}
Table~\ref{tab:metadata} specifies the failure-time context and audit fields for each case.

\subsection{Failure-Category Distribution}
\begin{table}[t]
\centering
\caption{Distribution of $128$ failure cases across root-cause~categories.}
\label{tab:rc_distribution}
\setlength{\tabcolsep}{4pt}
\resizebox{\columnwidth}{!}{%
\begin{tabular}{l l c c}
\toprule
\textbf{Family} & \textbf{Category} & \textbf{\#} & \textbf{\%} \\
\midrule
Source code             & RC1 (Compilation errors)           & 83 & 64.84 \\
Project dependency      & RC2 (Dependency issues)            & 29 & 22.66 \\
Build and configuration & RC3 (Config/build file issues)     & 6  & 4.69 \\
Source code             & RC4 (Missing source files)         & 4  & 3.12 \\
External resources      & RC5 (Download issues)              & 2  & 1.56 \\
Environment             & RC6 (Hardware issues)              & 1  & 0.78 \\
Command and arguments   & RC7 (Command/argument issues)      & 1  & 0.78 \\
Fuzz target             & RC8 (Broken fuzz target)           & 1  & 0.78 \\
Miscellaneous           & RC9 (Fuzzer regression)            & 1  & 0.78 \\
\bottomrule
\end{tabular}%
}
\end{table}

\subsection{Benchmark Selection}
The evaluation ledger counts archived build-failure logs. Confirmed same-origin failures are grouped before metadata checks. Matching error text alone does not establish a duplicate. Duplicate groups are defined by confirmed underlying failures, rather than matching error text alone. After deduplication, the metadata/accessibility check requires extractable logs, failure-time commits and configuration, and initially resolvable resource references. No historical patch is required. Passing this check does not guarantee executable restoration. The final stage pins the OSS-Fuzz/upstream commits, image digest, and dependency commit identifiers, restores the Docker environment, and requires a matching fatal diagnostic and failure stage, manually confirmed. It excludes unavailable execution-time resources, failed historical restoration, and nonmatching replay. The screening process excludes 1,076 duplicates, 256 records with incomplete metadata or inaccessible resources, and 360 records that fail historical restoration or reproduction checks. These sum to 1,692, leaving 128 cases from 113 projects. These stage totals are not estimates of the prevalence of failure causes in OSS-Fuzz.

\subsection{Strict FGRK Source Isolation}\label{app:fgrk}
Table~\ref{tab:fgrk_flow} summarizes the sequential source-selection process. The FGRK ledger separately counts historical failure-and-repair candidate materials. Multiple reports, PRs, or commits may initially refer to one incident. The source window is December 29, 2022 to June 14, 2026. Both failure evidence and the repair information actually extracted must satisfy this window. For evolving pages, the used content version is preserved and checked. Additions after June 14, 2026 and unverifiable content versions are excluded. The shared collection window does not imply identical candidate ledgers or independence between them.

Among 210 normalized source projects, 86 match the exclusion list formed from all 113 evaluation projects, so their entire 253 records are removed. This is broader than excluding matching evaluation case identifiers. Fork, mirror, shared-upstream, and patch-propagation checks then remove all 42 records of 14 related projects plus seven same-origin incidents from other retained projects, totaling 49 exclusions. The last 12 traceability exclusions do not eliminate any additional project. Common error types and general build principles are not themselves exclusion grounds.

Only the final 188 failures from 110 projects are used to regenerate templates. Previously mixed-source templates are not deemed isolated merely by deleting project names. Content depending on excluded sources is re-extracted from permitted material. Two authors independently review technical correctness, provenance, and case specificity. The resulting 43 ERR entries contain diagnostic patterns, applicability/exclusion conditions, repair actions, and validation requirements. The 13 GL entries contain reusable build principles and operational constraints. The 56-entry total and a maximum of four combined ERR/GL entries per prompt are distinct quantities.

Runtime queries use FAILURE\_REGION when available and a fixed log-tail slice otherwise. Positive and negative regular expressions select ERRs, and keywords associate GLs. Fixed priority, deduplication, and tie-breaking rules limit injection to four combined entries. Reference fixes and source audits are separate from the Solver's snapshots, mounts, and tools. Memory and related caches reset between cases and runs. The exact tie-breaking rule and log-tail length remain unspecified. The reported experiments use this strict knowledge-base version.


\end{document}